# Isotope coefficient of optimally doped high-Tc cuprates

A. Messad

*Laboratoire de Physique, Lycée Olympe de Gouges, 93130 Noisy-Le-Sec, France*

*Within the framework of pure BCS (i.e. when the critical temperature is proportional to the Debye temperature), we show that the isotope coefficient is always less than ½ and could be extremely small in polyatomic superconductors, depending on the chemical formula unit. This finding leads to quantitative explanation of observed values (correct orders of magnitude and sometimes exact numerical values) in optimally doped cuprates, magnesium diboride and alkali-doped fullerenes. Consequently, weakness of the isotope coefficient is not only compatible with pure electron-phonon interaction, but this is perhaps the rule in polyatomic systems.*

Two questions arise when one examines the isotope effect in the cuprates [1]:

(a) Why is the isotope coefficient $\alpha$ small in optimally doped systems?

(b) Why does $\alpha$ increase rapidly by further under(over)doping of the system?

Possible answers can be found if one considers the combined effect of a polyatomic system, the coulomb interaction and the band structure [2].

In this paper, we shall derive, within the framework of pure BCS, a formula for $\alpha$ dealing with the polyatomic effect and apply it to the cuprates to answer the first question. We shall also test this formula on well-established BCS-systems [3] like $MgB_2$, $K_3C_{60}$ and $Rb_3C_{60}$.

It was observed by [4], that the critical temperature of a polyatomic superconductor is given by

$$T_c \propto M^{-1/2}, \qquad (1)$$

where $M$ is the molar mass (i.e. the mass of the chemical formula unit).

Suppose now that a chemical element $X$ is substituted by its isotope $X^*$. By differentiating eq.(1) with respect to $M_X$ (the atomic mass of $X$), we obtain

$$\frac{\Delta T_c}{T_c} = -\frac{1}{2}\frac{xM_X}{M}\frac{\Delta M_X}{M_X}, \qquad (2)$$

where $\Delta M_X = M_X - M_{X^*}$ and $x$ is the stoichiometric coefficient of $X$ in the chemical formula unit.

Classically, the isotope coefficient is given by

$$\frac{\Delta T_c}{T_c} = -\alpha(X)\frac{\Delta M_X}{M_X} \qquad (3)$$

Equation (3) is used to calculate $\alpha(X)$ from experimental values of $\Delta T_c$ and $T_c$.

By comparing (2) and (3), we can identify the isotope coefficient as

$$\alpha(X) = \frac{1}{2}\frac{xM_X}{M} \qquad (4)$$

And when the substitution is incomplete and effected with a rate r, we have

$$\alpha(X) = \frac{1}{2}\frac{rxM_X}{M} \qquad (5)$$

Exact calculation gives

$$\alpha(X) = \frac{M_X}{\Delta M_X}\left[\left(1 - \frac{xr\Delta M_X}{M}\right)^{-\frac{1}{2}} - 1\right] \qquad (6)$$

Note that eq.(5) is an approximation of eq.(6) for small $(\Delta M_X / M)$ and then for large chemical formula units as in the cuprates.

According to eq.(5), $\alpha$ is always inferior to ½ for polyatomic superconductors and equal to ½, the canonical value given by the standard BCS theory, for monoatomic ones (i.e. $X_x$).

Comparison between calculated values (eq. (5)) and experimental data is shown in the following table.

| Superconductor | $M$ (g/mol) | Ref. | $T_c$ (k) | $X$ | $r$ | Experimental $\alpha_{(X)}$ | eq. (5) $\alpha_{(X)}$ (*) |
|---|---|---|---|---|---|---|---|
| $La_{1.85}Sr_{0.15}CuO_4$ | 397.6 | [1] [2] [5] [1] | 35 ? 37.6 optimal | O Cu | ? ? 1 ? | 0.1-0.2 0.07 0.08 $\alpha(Cu) \approx \alpha(O)$ | 0.080 0.080 |
| $YBa_2Cu_3O_7$ | 665.7 | [1] [5] [1] | 91 91.4 60 (plateau) | O Cu | ? 1 ? | 0.02-0.05 0.024 -0.14-(-0.34) | 0.084 0.143 |
| $YBa_2Cu_4O_8$ | 745.5 | [5] [6] | 80.8 81.6 | O | 1 0.95 | 0.048 0.0805 | 0.086 0.081 |
| $Bi_2Sr_2CaCu_2O_8$ | 888.3 | [1] [7] | 76 92 | O | ? ? | 0.03 0.087 | 0.072 |
| $Bi_2Sr_2Ca_2Cu_3O_{10}$ | 1055.9 | [1] | 110 | O | ? | 0.03-(-0.013) | 0.076 |
| $Nd_{1.85}Ce_{0.15}CuO_4$ | 415.3 | [1] [8] | 24 24 | O | 1? 0.85 | <0.05 ≤0.05 | 0.077 0.065 |
| $HoBa_2Cu_4O_8$ | 821.5 | [9] [10] | 79 79 | O Cu | ? ? | 0.05 0.214 | 0.078 0.154 |
| $MgB_2$ | 45.9 | [11] [12] [13] | 39 | B | ? | 0.26 0.28 0.30 | 0.24 |
| $Rb_3C_{60}$ | 976.5 | [14] [15] | 19 ? | C | 0.75 0.99 | 0.37 0.21 | 0.28 0.38 |
| $K_3C_{60}$ | 837.3 | [2] | ? | C | ? | 0.37 or 1.74 | 0.45 |

(*) We assume r = 1 when experimental value is not available.

The calculated isotope coefficient has the correct order of magnitude in general and some exact numerical values at optimal doping (maximum of $T_C$). This fact suggests that in the latter case, superconductivity is perhaps caused by electron-phonon interaction alone. Negative values cannot be explained by the polyatomic effect alone.
For optimal La-Sr-Cu-O system it is natural to have the same isotope coefficient for both copper and oxygen, because $M_{Cu} \approx 4M_O$. The same thing is expected for optimal Nd-Ce-Cu-O system.
Tested on well-established BCS systems (last three compounds of the table), eq.(5) gives fairly accurate results.
Equation (5) can be used to reveal the presence of electron-phonon interaction as follows. If α calculated for one or several chemical elements is equal to (or not far from) the observed one, then we can consider that this interaction exists and these elements are involved in it via their vibration modes. If not, other effects should be taken into account or there is another kind of interaction behind the superconductivity.

To summarize, we have established a BCS criterion for polyatomic superconductors, which is not unique as in monoatomic ones but depends on the chemical composition. This criterion teaches us that in such superconductors, the isotope effect could be very small or even negligible and still perfectly compatible with pure electron-phonon interaction.

**Acknowledgements:** I thank F. Pillier for the bibliography and M. Herbaut for correcting the English.